\documentclass[a4paper,11pt]{article}
\pdfoutput=1

\usepackage{jheppub}

\usepackage[T1]{fontenc} 
\usepackage{graphicx}
\usepackage{epsfig}
\usepackage{subcaption}
\usepackage[titletoc,title]{appendix}
\usepackage{mathtools}
\usepackage[percent]{overpic}

\title{\boldmath  Lifshitz quasinormal modes and relaxation \\from holography}

\author{Watse Sybesma and}
\author{Stefan Vandoren}

\affiliation{
Institute for Theoretical Physics and Center for Extreme Matter and Emergent Phenomena,\\Utrecht University, 3584 CE, Utrecht, The Netherlands.}

\emailAdd{Z.W.Sybesma@uu.nl}
\emailAdd{S.J.G.Vandoren@uu.nl}

\abstract{\\
We obtain relaxation times for field theories with Lifshitz scaling and with holographic duals Einstein-Maxwell-Dilaton gravity theories. This is done by computing quasinormal modes of a bulk scalar field in the presence of Lifshitz black branes. We determine the relation between relaxation time and dynamical exponent $z$\,, for various values of boundary dimension $d$ and operator scaling dimension. It is found that for $d>z+1$\,, at zero momenta, the modes are non-overdamped, whereas for $d\leq z+1$ the system is always overdamped. For $d=z+1$ and zero momenta, we present analytical results.
}

\begin{document} 
\maketitle
\flushbottom

\section{Introduction}
\label{sec:intro}

Much effort has been invested into the understanding of the application of holography since its seminal papers \cite{Maldacena:1997re,Witten:1998qj,Gubser:1998bc}. Motivated by the fact that real world systems often exhibit non-relativistic scale invariance at critical points, rather than relativistic invariance, it is of interest to extend the holographic dictionary into non-relativistic context \cite{Hartnoll:2009aa,Kachru:2008yh,Balasubramanian:2009rx,Son:2008ye,Taylor:2008tg}. As a model of non-relativistic scaling at a critical point on the boundary we require invariance under
\begin{equation}\label{lifscaling}
	t
	\rightarrow
	\lambda^{z}
	t
	\,,
	\qquad
	\;\;
	\vec{x}
	\rightarrow
	\lambda 
	\vec{x}
	\,,
\end{equation}
where $t$ denotes time, $\vec{x}$ denotes spatial coordinates on the holographic boundary and $z$ is called the dynamical exponent. For $z=1$ the theory is Lorentz invariant. When $z>1$ the system obtains anisotropic scaling between space and time, often called Lifshitz scaling, which violates Lorentz boost invariance.

The main goal of this paper is to determine the relaxation time of the $d$-dimensional boundary theory with dynamical exponent $z$ at some temperature $T$. Relaxation occurs after having perturbed the system by an operator at the boundary. We consider operators with scaling dimension $\Delta$ and a spin-zero field as holographic dual in the bulk. We obtain the relaxation time by computing quasinormal modes in the bulk and investigate the dependence on $d$\,, $z$ and $\Delta$\,.

To obtain Lifshitz scaling and temperature on the boundary, we consider a black brane solution of an Einstein-Maxwell-Dilaton (EMD) action \cite{Taylor:2008tg,Tarrio:2011aa}. A black brane in the bulk drives field excitation into dissipation. We then solve the (complexified) Klein-Gordon equation of the spin-zero field in the probe limit. The resulting complex eigenvalues are called the quasinormal frequencies corresponding to quasinormal modes. The smallest imaginary part of these eigenvalues is inversely proportional to the relaxation time $\tau$ of the boundary system. 

The quasinormal modes for $z=1$ with $d=3\,,\,4\,,\,6$ were studied numerically for the first time in \cite{Horowitz:2000aa}. An analytic solution for $d=2$\,, $z=1$ was obtained in \cite{Birmingham:2001pj}. In this current paper we find a generalization of this analytic solution for $d=z+1$ for vanishing momenta. It reads
\begin{equation}
	\omega_{n}
	=
	-i
	2\pi T
	\left(
		2n+\frac{\Delta}{z}
	\right)
	,
	\;\;
	\Rightarrow
	\;\;
	\tau
	=
	\frac{z}{2\pi T\Delta}
	\,,
\end{equation}
where $T$ denotes temperature and $n=0$ gives the lowest lying quasinormal mode. For $d=3$, $z=2$ this equation coincides with the one found in \cite{Myung:2012cb}. For reviews on this topic we recommend \cite{Berti:2009kk,Konoplya:2011qq}.

A summary of previously obtained analytic and numerical solutions of quasinormal modes of spin-zero fields, making usage of various bulk actions, entails: $d=2$\,, $z=3$ in New Massive Gravity (NMG) \cite{CuadrosMelgar:2011up}, $d\geq4$\,, $z=2$ in a $R^{2}$ gravity setting \cite{Abdalla:2012si} and $d\geq2$\,, $z=2$ in a $R^{3}$ gravity setting \cite{Giacomini:2012aa}. The case for $d=3$\,, $z=2$ has been studied in the Einstein-Proca-Scalar (EPS) background \cite{Gonzalez:2012de,Balasubramanian:2009rx}, in the EMD setup \cite{Myung:2012cb} and in a topological black hole in a Einstein-Maxwell-Proca (EMP) background \cite{Gonzalez:2012xc,Mann:2009yx,Brynjolfsson:2009ct}. All these quasinormal mode solutions were found to be purely imaginary. This signals that the corresponding system is overdamped. In the papers \cite{Myung:2012cb,Abdalla:2012si} it was therefore conjectured that for (most) Lifshitz black holes the quasinormal modes are purely imaginary. Our conclusions will be different. In our numerical analysis we find that for $d>z+1$ the quasinormal modes have a real component. However, for the case of $d\leq z+1$ one continues to find overdamped solutions.

A brief outline of the paper is the following. In the next section we introduce notations and derive the Schr\"{o}dinger-like equation for quasinormal modes of a spin-zero field. In Section $3$ we obtain the new analytic solution for quasinormal modes and analyze the remaining cases using numerics. Here we chart the (non-)overdamped region. Next, in Section 4 we present the relaxation times and their behavior versus $z\,$, $d\,$ and $\Delta$. Finally, in Section 5 we present an outlook for possible future work.
\section{Schr\"{o}dinger problems for Lifshitz geometries}
The objective is to obtain relaxation times for a boundary field theory probed by an operator dual to a spin-zero field in the bulk. In order to pursue this goal we need to compute quasinormal modes of a spin-zero field in an asymptotically Lifshitz black brane background. This requires solving the equation of motion of a massive scalar field in the EMD background, in the probe limit and subjected to appropriate boundary conditions.
\subsection{Lifshitz brane}
The metric line element of a black brane exhibiting Lifshitz scaling (\ref{lifscaling}) on the boundary can be expressed as
\begin{equation}\label{metric}
	ds^{2}
	=
	\frac{1}{r^{2}V^{2}(r)}dr^{2}
	-
	V^{2}(r)r^{2z}dt^{2}
	+
	r^{2}
	d\vec{x}^{2}_{d-1}
	\,,
\end{equation}
where $r$ is the radial bulk coordinate, which under Lifshitz symmetry scales as $r\rightarrow\lambda r$\,. The limit $r\rightarrow\infty$ corresponds to the boundary. The Lifshitz radius, a generalization of the Anti-de Sitter radius, is put to unity in this paper. The blackening factor in the EMD setup\footnote{For different bulk fields, such as e.g. the EPS setup in $d=3$ and $z=2$, one has \cite{Gonzalez:2012de,Balasubramanian:2009rx}
\begin{equation}
	V^{2}_{\text{EPS}}=1-\frac{r^{2}_{h}}{r^{2}}
	\,.
\end{equation}} is \cite{Taylor:2008tg}
\begin{equation}
	V^{2}
	=
	1
	-
	\left(\frac{r_{h}}{r}\right)^{d+z-1}
	,
\end{equation}
where $r_{h}$ denotes the horizon. The temperature of the black brane is given by
\begin{equation}
	4\pi T
	=
	(d+z-1)r_{h}^{z}
	\,.
\end{equation}
To find the quasinormal modes we need to compute the tortoise coordinate $r_{*}$\,, which characterizes the radial null curves obeying $t=\pm r_{*}+\text{constant}$. We consequently demand
\begin{equation}
	dr_{*}
	=
	\frac{1}{r^{z+1}V^{2}}
	dr
	\,.
\end{equation}
The general solution, for $d+z>1$\,, is
\begin{equation}
	r_{*}=-\frac{r^{-z}}{z}\left.\right._{2}F_{1}\left[1,\frac{z}{d+z-1};1+\frac{z}{d+z-1};\left(\frac{r_{h}}{r}\right)^{d+z-1}\right]
	,
\end{equation}
in terms of the hypergeometric function of the second kind. Without any loss of generality we fix the integration constant to be zero. For the special case of $d=z+1$ it becomes
\begin{equation}\label{hypergauss}
	\left.r_{*}\right|_{d=z+1}=-\frac{1}{2zr_{h}^{z}}\log
		\left[
			\frac{1+\frac{r^{z}_{h}}{r^{z}}}		
			{1-\frac{r^{z}_{h}}{r^{z}}}
		\right]		
	.
\end{equation}
Next we define Eddington-Finkelstein coordinates $v$ and $u$ as
\begin{equation}
	v
	=
	t
	+
	r_{*}
	\,,
	\;\;
	\qquad
	u
	=
	t
	-
	r_{*}
	\,,
\end{equation}
and we require infalling boundary conditions, for a field $\phi$ near the horizon to be
\begin{equation}\label{horizoncondition}
	\phi(r\rightarrow r_{h})
	\sim
	e^{-i\omega v}
	=
	e^{-i\omega t-i\omega r_{*}}
	.
\end{equation}
This condition causes dissipation and complexifies the field $\phi$\,.
\subsection{Quasinormal modes of a scalar probe}
We consider the equation of motion of a massive scalar field $\phi$ on the background of (\ref{metric})
\begin{equation}\label{kg}
	\Box\phi
	=
	m^{2}\phi
	\,.
\end{equation}
Assuming the probe limit we ignore back reaction on the metric. We require boundary conditions such that at the horizon the field $\phi$ has to be falling into the black brane and that at radial infinity the field goes to zero, $\phi(r\rightarrow\infty)=0$\,. This system has a dissipative nature since over time more and more field excitations will fall into the black brane while the excitations at radial infinity are reflected. There is no incoming flux from the horizon nor from the boundary. Quasinormal modes are the discrete values of energy $\omega$ for which the equation of motion of $\phi$ is satisfied, taking into account these boundary conditions. Making a plane wave Ansatz and a radial rescaling
\begin{equation}\label{planaransatz}
	\phi
	=
	r^{\frac{d-1}{2}}\tilde{\phi}(r)e^{-i\omega t+i \vec{k}\cdot\vec{x}}
	\,,
\end{equation}
where $\vec{k}$ denotes momentum, the Klein-Gordon equation in (\ref{kg}) can be written as a one-dimensional time independent Schr\"{o}dinger-like equation 
\begin{equation}\label{schroedinger}
	\left[
		\partial_{*}^{2}
		+
		\omega^{2}
		-
		\mathbb{V}(r)
	\right]
	\tilde{\phi}
	=
	0
	\,,
	\;\;
	\qquad
	\partial_{*}
	=
	r^{z+1}V^{2}\partial_{r}
	\,.
\end{equation}
The corresponding potential is
\begin{equation}
	\mathbb{V}(r)
	=
	r^{2z}V^{2}(r)
	\left(
		\frac{k^{2}}{r^{2}}
		+
		m^{2}
		+
		\frac{(d-1)(d+2z-1)}{4}
		+
		\frac{(d-1)^{2}}{4}
		\left(
			\frac{r_{h}}{r}
		\right)^{d+z-1}
	\right)
	.
\end{equation}
Asymptotically, near the boundary, the solution to the Schr\"{o}dinger-like equation is given by
\begin{equation}\label{radialinfty}
	\phi(r)
	\sim
	Ar^{-\Delta_{-}}
	+
	Br^{-\Delta_{+}}
	,
	\;\;
	\Delta_{\pm}
	=
	\frac{d+z-1}{2}
	\pm
	\sqrt{\left(\frac{d+z-1}{2}\right)^{2}+m^{2}}
	\,,
\end{equation}
where $A$ and $B$ are independent of $r$\,. Applying the regular recipe for holography \cite{Witten:1998qj,Gubser:1998bc}, we consider the term containing $\Delta\equiv\Delta_{+}$ to be the normalizable mode and hence the scaling dimension of the operator $\mathcal{O}_{\Delta}$ dual to $\phi$ is $\Delta$. Considering the Klein-Gordon inner product we find that for this mode to be normalizable we have to constrain the scalar mass \cite{Andrade:2012xy}
\begin{equation}\label{bfbound1}
	\int^{\infty}_{r_{h}}
	dr
	r^{d-z-2}|Br^{-\Delta}|^{2}
	<
	\infty
	\,,
	\;\;
	\Rightarrow
	\;\;
	m^{2}
	>
	-
	\left(
		\frac{d+z-1}{2}
	\right)^{2}
	\equiv
	m^{2}_{BF}
	\,,
\end{equation}
where the contribution $r^{d-z-2}$ is due to a volume factor resulting from taking a space-like slice as integration area. The $m^{2}_{BF}$ is the minimal value above which $m^{2}$ has to remain. This is the Breitenlohner-Freedman (BF) bound. We require 
\begin{equation}\label{inftycondition}
	A(\omega,k,\Delta,d,z,T)=0
	\,,
\end{equation}
in order to have no incoming or outgoing flux at radial infinity. 

\begin{figure}[h]
	\vspace{0.5cm}
        \centering
        \begin{subfigure}[b]{.60\textwidth}
                \begin{overpic}[width=1\textwidth]{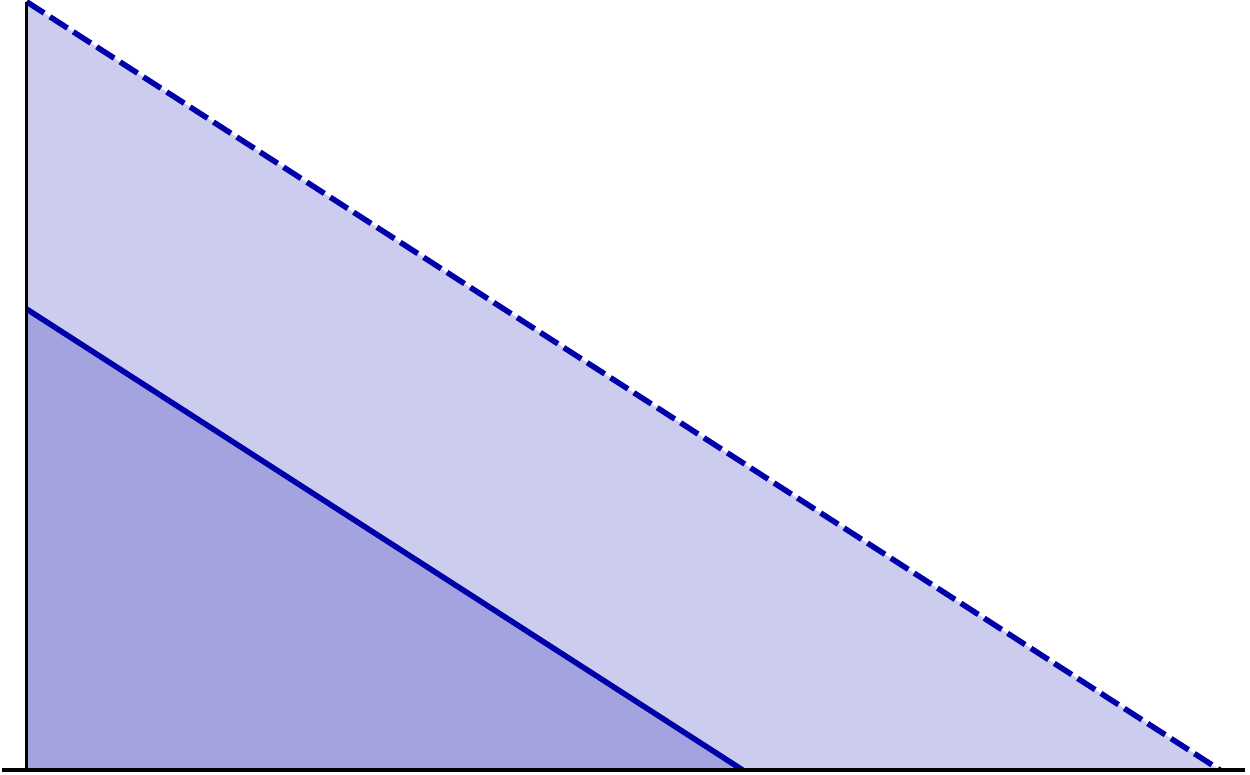}

		\put (-3,0) {$\displaystyle 0$}
 		\put (+1.64,62.5) {|}
		\put (+1.53,62.5) {|}
		\put (98.5,-0.18){\rotatebox{90}{|}}
		\put (98.5,-0.28853){\rotatebox{90}{|}}
 
		\put (-3,59) {$d$}
		\put (100,1) {$\displaystyle z$}
		\put (58,43) {$\displaystyle \text{violates BF}$}
		\put (43,18) {$\displaystyle \text{relevant}$}
		\put (15,9) {$\displaystyle \text{irrelevant}$}
		\put (2,38) {\rotatebox{-32}{$\displaystyle d=1-z+\Delta$}}
		\put (13,56){\rotatebox{-32}{$\displaystyle d=1-z+2\Delta$}}
		\end{overpic}
        \end{subfigure}
        \caption{Visualization of (\ref{handigeregel}). Exactly at the $d=1-z+\Delta$ line the operator is marginal. On and above the line $d=1-z+2\Delta$ the BF bound is violated.}\label{figure0}
\end{figure}
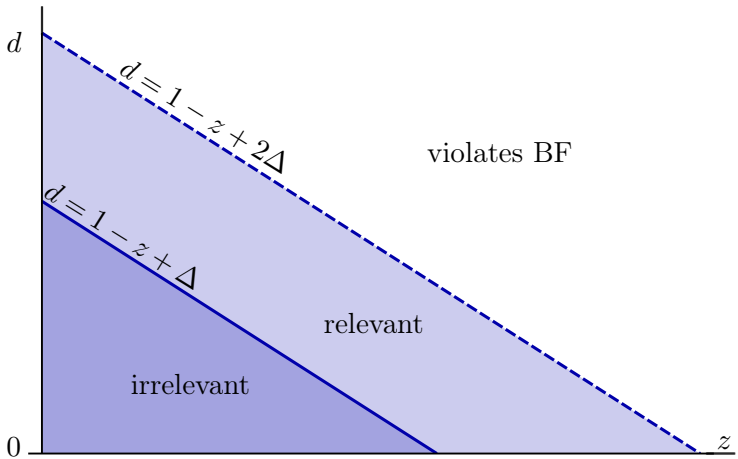

The BF bound (\ref{bfbound1}) translates into the following bound on the scaling dimension
\begin{equation}\label{bfbound}
	\Delta
	=
	\frac{d+z-1}{2}
	+
	\sqrt{\left(\frac{d+z-1}{2}\right)^{2}+m^{2}}
	>
	\frac{d+z-1}{2}
	\,.
\end{equation}
Marginal operators in a theory with Lifshitz scaling have $\Delta_{\text{marginal}}=d+z-1$\,, so we summarize
\begin{equation}\label{handigeregel}
	\frac{d+z-1}{2}
	<
	\Delta_{\text{relevant}}
	<
	\Delta_{\text{marginal}}
	=
	d+z-1
	<
	\Delta_{\text{irrelevant}}
	\,,
\end{equation}
which can be visualized as in Figure \ref{figure0}.
A quasinormal mode is defined by solving the Schr\"{o}dinger-like equation (\ref{schroedinger}) supplemented with boundary condition (\ref{horizoncondition}) at the horizon and boundary condition (\ref{inftycondition}) at radial infinity. Quasinormal modes $\omega_{n}$ are then obtained from
\begin{equation}\label{defqnms}
	A(\omega,k,\Delta,d,z,T)
	=
	0
	\,,
	\;\;
	\Rightarrow
	\;\;
	\omega_{n}
	=
	\omega_{\text{Re}}(n,k,\Delta,d,z,T)
	-
	i
	\omega_{\text{Im}}(n,k,\Delta,d,z,T)
	\,,
\end{equation}
where integer $n$ labels the overtone number of the quasinormal mode. By definition the $n=0$ gives the imaginary component closest to zero. 

When $\omega_{\text{Im}}$ is positive (in our conventions) it implies stability of the gravitational background under scalar perturbations. We can verify that $\omega_{\text{Im}}$ is always positive due to the fact that the potential is real, positive and strictly increasing. This conclusion follows from a reasoning similar to the one given in \cite{Horowitz:2000aa}, but adapted to Lifshitz scaling. One starts by introducing $\tilde{\phi}=e^{-i\omega r_{*}}\hat{\phi}(r)$ to (\ref{schroedinger}) in order to obtain
\begin{equation}
	\left[
	r^{z+1}V^{2}(r)
	\partial_{r}^{2}
	+
	\left[
		\partial_{r}
		\left(
			r^{z+1}V^{2}(r)
		\right)
		-
		2i\omega
		\right]
		\partial_{r}
		-
		\frac{\mathbb{V}(r)}{r^{z+1}V^{2}}
	\right]
	\hat{\phi}
	=
	0
	\,.
\end{equation}
Multiplying this equation by $\hat{\phi}^{*}$ and integrating over $r$ from $r_{h}$ to $\infty$\,, after integration by parts, yields
\begin{equation}\label{hheq}
	\int^{\infty}_{r_{h}}
	dr
	\left[
		r^{z+1}
		V^{2}(r)|\partial_{r}\hat{\phi}|^{2}
		+
		2i\omega \hat{\phi}^{*}\partial_{r} \hat{\phi}
		+
		\frac{\mathbb{V}(r)}{r^{z+1}V^{2}(r)}
		|\hat{\phi}|^{2}
	\right]
	=
	0
	\,.
\end{equation}
Taking the imaginary part of (\ref{hheq}), applying integration by parts and inserting the result back into (\ref{hheq}) results in\begin{equation}
	\int^{\infty}_{rh}
	dr
	\left[
		r^{z+1}
		V^{2}(r)
		|
			\partial_{r}\hat{\phi}
		|^{2}
		+
		\frac{\mathbb{V}(r)}{r^{z+1}V^{2}(r)}
		|
		\hat{\phi}
		|^{2}
	\right]
	=
	-
	\frac{|\omega|^{2}|\hat{\phi}(r_{h})|^{2}}{\text{Im}\omega}
	\,,
\end{equation}
where the left-hand side is ensured to remain positive with the potential under consideration in this paper. This guarantees a negative imaginary value of $\omega$.

The connection between relaxation time and quasinormal modes follows from
\begin{equation}
	|e^{-i \omega_{n}t}|
	=
	|e^{-i \omega_{\text{Re}}t}e^{- \omega_{\text{Im}}t}|	
	=
	e^{- \omega_{\text{Im}}t}
	=
	e^{- t/\tau}
	\,,
\end{equation} 
which violates unitarity due to dissipation. The relaxation time $\tau$ is defined as 
\begin{equation}\label{thetau}
	\tau\equiv 1/\omega_{\text{Im}}
	\,.
\end{equation}
Notice that the real part of the quasinormal mode is referred to as the dispersion relation. For purely imaginary quasinormal modes we thus speak of the corresponding system being overdamped.
\section{Obtaining quasinormal modes}
Finding the quasinormal modes by solving the Schr\"{o}dinger-like equation (\ref{schroedinger}) supplemented with boundary condition (\ref{horizoncondition}) at the horizon boundary condition (\ref{defqnms}) at radial infinity is mostly done numerically in this paper. The case $d=z+1$, however, can be treated analytically for vanishing momenta, so we can consider this case separately.
\subsection{Analytic solutions, $d=z+1$}
For the EMD background, an analytical solution of quasinormal modes is known $d=3$, $z=2$ with $k=0$ \cite{Myung:2012cb}. We now determine additional analytical solutions for the general case of $d=z+1$ with $k=0$\,. 

Why this case is special from an analytic point of view can be understood from observing two simplifications which occur to (\ref{schroedinger}) in this case. Firstly, the potential only depends on terms containing $r^{2z}$. This simplifies the $r$-dependence of the potential $\mathbb{V}$ tremendously. Secondly, $r_{*}$ depends only on $r^{z}$ and can be analytically inverted to $r^{z}(r_{*})$\,, which enables one to directly express the potential $\mathbb{V}$ in terms of $r_{*}$\,. In other cases one is unable to accomplish this or it simply yields a much more complex expression.

However, in order to compute this analytic expression it is convenient to switch back from $\tilde{\phi}$ to $\phi$\,, require $d=z+1$\,, and write (\ref{schroedinger}) in the form
\begin{equation}
	\left(
		-r^{2z}V^{2}(r)m^2+\omega^{2}-r^{2z}V^{2}(r)\frac{k^2}{r^2}
	\right)\phi
	+
	\partial_{*}^{2}\phi
	+
	r^{z+1}V^{2}(r)
	\frac{z}{r}\partial_{*}\phi
	=
	0
	\,.
\end{equation}
We make use of the substitution 
\begin{equation}
	y
	=
	\left(
	\frac{r_{h}}{r}	
	\right)^{2}
	\,,
\end{equation}
in order to obtain
\begin{equation}
	\begin{aligned}
		\frac{
				\omega^{2}y^{z}-r_{h}^{2z-2}\left[m^{2}r_{h}^{2}+k^{2}y\right](1-y^{z})
			}
			{
				r_{h}^{2z}(1-y^{z})^{2}y
			}\phi
		-
			4\frac{
				(z-1)
				+
				y^{z}
			}
			{
				1-y^{z}
			}
			\phi'
		+
			4y\phi''
		=
		0
		\,,
	\end{aligned}
\end{equation}	
which, when assuming $k=0$\,, is solved by
\begin{equation}
	\begin{aligned}
		\phi
		=
		&
		\,c_{1}
		y^{\frac{\Delta_{-}}{2}}
		\left(
			1-y^{z}
		\right)^{\beta}
		\left.\right._{2}F_{1}\left[
			\frac{\Delta_{-}}{2z}+\beta,
			\frac{\Delta_{-}}{2z}+\beta;
			\frac{\Delta_{-}}{z};
			y^{z}
		\right]
		\\
		&
		+
		c_{2}
		y^{\frac{\Delta_{+}}{2}}
		\left(
			1-y^{z}
		\right)^{\beta}
		\left.\right._{2}F_{1}\left[
			\frac{\Delta_{+}}{2z}+\beta,
			\frac{\Delta_{+}}{2z}+\beta;
			\frac{\Delta_{+}}{z};
			y^{z}
		\right]
		\,.
	\end{aligned}
\end{equation}	
In the last expression $c_{1}$, $c_{2}$ are some constants and 
\begin{equation}
	\beta
	=
	\frac{i\omega}{4\pi T}
	\,.
\end{equation}
The deltas are the same as given in (\ref{radialinfty}). We continue by requiring both boundary conditions. We start with the boundary condition at $r\rightarrow\infty$\,, which corresponds to $y\rightarrow0$\,. We want Dirichlet boundary conditions to hold here. Taking the limit one obtains
\begin{equation}
	\phi(y\rightarrow 0)
	\sim
	c_{1}y^{\frac{\Delta_{-}}{2}}
	+
	c_{2}y^{\frac{\Delta_{+}}{2}}
	\sim
	c_{1}r^{-\Delta_{-}}
	+
	c_{2}r^{-\Delta_{+}}
	\,,
\end{equation}
thus we put $c_{1}=0$ in order to match the required boundary behavior from (\ref{inftycondition}). When taking the limit we made use of the identity
\begin{equation}\label{handy}
	\left.\right._{2}F_{1}[a,b;c;0]
	=
	1
	\;\;
	\text{for all $a$, $b$, and $c$ non-zero}
	,
\end{equation}
which can be found in e.g. \cite{abramowitz}. Next we apply the boundary condition at the horizon $r=r_{h}$\,, which corresponds to $y=1$\,. We can relate the behavior at $y=1$ to the behavior at $y=0$ using
\begin{equation}
	\begin{aligned}
		\left.\right._{2}F_{1}[a,a;c;z]
		=
		&
		\,
		\frac{\Gamma(c)\Gamma(c-2a)}{\Gamma(c-a)^{2}}
		\left.\right._{2}F_{1}[a,a;2a-c+1;1-z]
		\\
		&
		+
		(1-z)^{c-2a}
		\frac{\Gamma(c)\Gamma(2a-c)}{\Gamma(a)^{2}}
		\left.\right._{2}F_{1}[c-a,c-a;c-2a+1;1-z]
		\,,
	\end{aligned}
\end{equation}
where $\Gamma$ denotes the gamma function. Using (\ref{handy}) we arrive at
\begin{equation}
	\begin{aligned}
		\phi
		=
		&
		\lim_{y\rightarrow1}
		\left(
			1-y^{z}
		\right)^{\beta}
		\left\{
			\frac{
			\Gamma\left(
				\Delta_{+}/z
			\right)
			\Gamma\left(
				-2\beta
			\right)
			}
			{
			\Gamma\left(
				\Delta_{+}/(2z)
				-\beta
			\right)^{2}
			}
			+
			(1-y^{z})^{-2\beta}
			\frac{
			\Gamma\left(
				\Delta_{+}/z
			\right)
			\Gamma\left(
				2\beta
			\right)
			}
			{
			\Gamma\left(
				\Delta_{+}/(2z)
				+\beta
			\right)^{2}
			}
		\right\}
		.
	\end{aligned}
\end{equation}
Now we have to identify the ingoing modes as written down in (\ref{horizoncondition}). From (\ref{hypergauss}) we collect
\begin{equation}
	\left.r_{*}\right|_{d=z+1}
	=
	\frac{1}{4\pi T}
	\left[\log\left(1-y^{z}\right)
	-2\log\left(1+y^{z}\right)
	\right]
	\,,
\end{equation}
such that we are able to identify
\begin{equation}
	e^{-i\omega t}
	\left(
		1-y^{z}
	\right)^{\pm\beta}
	=
	e^{-i\omega t}
	e^{\pm i\omega 
	\frac{
		\log\left(
			1-y^{z}
		\right)
	}
	{
		4\pi T
	}	
	}
	=
	e^{-i\omega\left(
		t
		\pm
		r_{*}
	\right)
	}
	(1+y^{z})^{\mp \frac{i\omega }{2\pi T}}
	\,,
\end{equation}
which leaves us to the conclusion that we have to put the term with the positive power of $\beta$ to zero. This is accomplished by requiring
\begin{equation}
	\frac{
			1
			} 
			{
			\Gamma\left(
				\Delta_{+}/(2z)
				-\beta
			\right)
			}
			=
			0
			\,,
			\;\;
			\Rightarrow
			\;\;
			\frac{\omega}{2\pi T}
			=
			-i
			\left(
			2n
			+
			\frac{\Delta}{z}
	\right)
	\,,
\end{equation}
where $n$ is a positive integer. For $z=2$ the result restores the finding in \cite{Myung:2012cb}. For $z=1$ we obtain the case of \cite{Birmingham:2001pj}. This solution holds for any $d>1$ and at all times exhibits overdamped behavior. It is concluded from (\ref{thetau}) that 
\begin{equation}
	\tau_{d=z+1}
	=
	\frac{z}{2\pi T \Delta}
	\,,
\end{equation}
which implies that higher anisotropy corresponds to a longer relaxation time. From the numerics it is observed that this holds as long as $z>d-1$\,.
\subsection{Numerical solutions, $d\neq z+1$}
Initially we started by solving the Schr\"{o}dinger-like equation (\ref{schroedinger}) using the method described in \cite{Horowitz:2000aa}. This method, in essence, solves the Schr\"{o}dinger-like equation by using a power series Ansatz. Applying the boundary conditions, results in a recursive relation between different terms in the power series Ansatz. This recursive relation is too cumbersome to be handled analytically. We thus have to resort to numerical methods. 

The time it takes for the power series to converge numerically, is found to increase dramatically for $z>1$\,. It was already noted by the authors of \cite{Horowitz:2000aa} themselves that increasing $d$ increases computation time. This is connected to the fact that increasing $d$\,, increases the power of $r$ in the potential, yielding a more complex recursive relation, which amplifies the time it takes to converge. Noting that when increasing $z$ that the power of $r$ in the potential gets larger too, explains the rise of computation time.

To decrease computational time we adopted the Improved Asymptotic Iteration Method, as described in \cite{Cho:2011sf}. This method, however making use of a recursive structure as well, relies on (as opposed to the previously mentioned algorithm) the observation that
\begin{equation}\label{aimer}
	\chi''(x)
	=
	\lambda_{0}(x)\chi'(x)
	+
	s_{0}(x)\chi(x)
	\,,
	\;\;
	\Rightarrow
	\;\;
	\chi^{(n+2)}(x)
	=
	\lambda_{n}(x)\chi'(x)
	+
	s_{n}(x)\chi(x)
	\,,
\end{equation}
where the superscripted $(n+2)$ denotes the order of derivation. The $\lambda_{0}$ and $s_{0}$ are polynomials to which $\lambda_{n}$ and $s_{n}$ are related in a recursive fashion including various orders of derivatives as well. From the ratio of the $(n+3)^{th}$ and $(n+2)^{th}$ derivatives, one shows
\begin{equation}\label{eigenfun}
	\frac{d}{dx}
	\log\left(
		\chi^{(n+2)}(x)
	\right)
	=
	\frac{\lambda_{n+1}\left(\chi'(x)+\frac{s_{n+1}(x)}{\lambda_{n+1}(x)}\chi(x)\right)}
	{\lambda_{n}\left(\chi'(x)+\frac{s_{n}(x)}{\lambda_{n}(x)}\chi(x)\right)}
	\,.
\end{equation}
Now we introduce the asymptotic aspect of the method. If for some sufficiently large $n$
\begin{equation}
	\frac{s_{n}}{\lambda_{n}}
	\approx
	\frac{s_{n+1}}{\lambda_{n+1}}
	\,,
\end{equation}
one can solve (\ref{eigenfun}) and, by plugging this back into (\ref{aimer}), find a solution for $\chi(x)$\,. This approach is called the Asymptotic Iteration Method and was originally developed by \cite{Cifti}. The \textit{Improved} Asymptotic Iteration Method entails, as modification to the original approach, some convenient power series expansions of $s_{n}$ and $\lambda_{n}$ in order to simplify their respective recursive relations.

To put this algorithm to our use we choose to rescale the coordinate $r$ into a dimensionless parameter with a finite range in the following way
\begin{equation}
	x
	=
	1-\frac{r_{h}}{r}
	\,.
\end{equation}
We rewrite equation (\ref{schroedinger}) into
\begin{equation}\label{difff}
		\tilde{\phi}''(x)
		+
		\underbrace{
		\left(
		\frac{
			\partial_{x}\left[(1-x)^{2}h(x)\right]
		}{
			(1-x)^{2}h(x)
		}
		\right)
		}_{\mathclap{\equiv Q(x)}}
		\tilde{\phi}'(x)
		+
		\underbrace{
		\left(
		r_{h}^{2} \frac{\omega^{2}-\mathbb{V}(x)}{\left[(1-x)^{2}h(x)\right]^{2}}
		\right)
		}_{\mathclap{\equiv R(x)}}
		\tilde{\phi}(x)
		=
		0
		\,,
\end{equation}
where $h(r)=r^{z+1}V^{2}(r)$ and the accents denote derivatives with respect to $x$\,.
We aim to scale out the behavior near the horizon and the boundary. This is done by employing the scaling
\begin{equation}
	\tilde{\phi}(x)
	=
	x^{A}(1-x)^{B}\chi(x)
	\,,
	\;\;
	A
	=
	\frac{-i\omega}{(d+z-1)r_{h}^{z}}
	\,,
	\;\;
	B
	=
	\frac{z}{2}
	+
	\sqrt{m^{2}+\frac{(d+z-1)^{2}}{4}}
	\,.
\end{equation}
Using that
\begin{equation}
	\partial_{x}
	\left[
		x^{A}(1-x)^{B}
	\right]
	=
	x^{A}(1-x)^{B}
	\overbrace{\left[
		Ax^{-1}
		-
		B(1-x)^{-1}
	\right]}^{\equiv F(x)}
	\,,
\end{equation}
we rewrite (\ref{difff}) as
\begin{equation}
	\chi''(x)
	+
	\underbrace{
	\left[
		2F(x)+Q(x)
	\right]}_{=\lambda_{0}}
	\chi'(x)
	+
	\underbrace{\left[
		F^{2}(x)
		+
		\partial_{x}F(x)
		+
		Q(x)F(x)
		+
		R(x)
	\right]}_{=s_{0}}\chi(x)
	=
	0
	\,.
\end{equation}
Now that the form of (\ref{aimer}) is obtained, we can in principle construct $\lambda_{n}$ and $s_{n}$ for any $n$\,. The quasinormal modes $\omega$ are obtained by requiring and solving
\begin{equation}
	s_{n}(x,\omega)\lambda_{n+1}(x,\omega)
	-
	s_{n+1}(x,\omega)\lambda_{n}(x,\omega)
	\approx
	0
	\,,
\end{equation}
for some large value of $n$\,. Typically we use $n\sim30$ to obtain certainty up to at least two decimals. In our case for $z>1$ we obtained quicker convergence with this algorithm, than when using the first mentioned algorithm. We reproduced the results from \cite{Horowitz:2000aa} for the case of a black brane in Table \ref{table1} in the appendix as a check.
\subsection{Analysis of numerical output}
\begin{figure}[h]
        \centering
        \begin{subfigure}[b]{.60\textwidth}
                \begin{overpic}[width=1\textwidth]{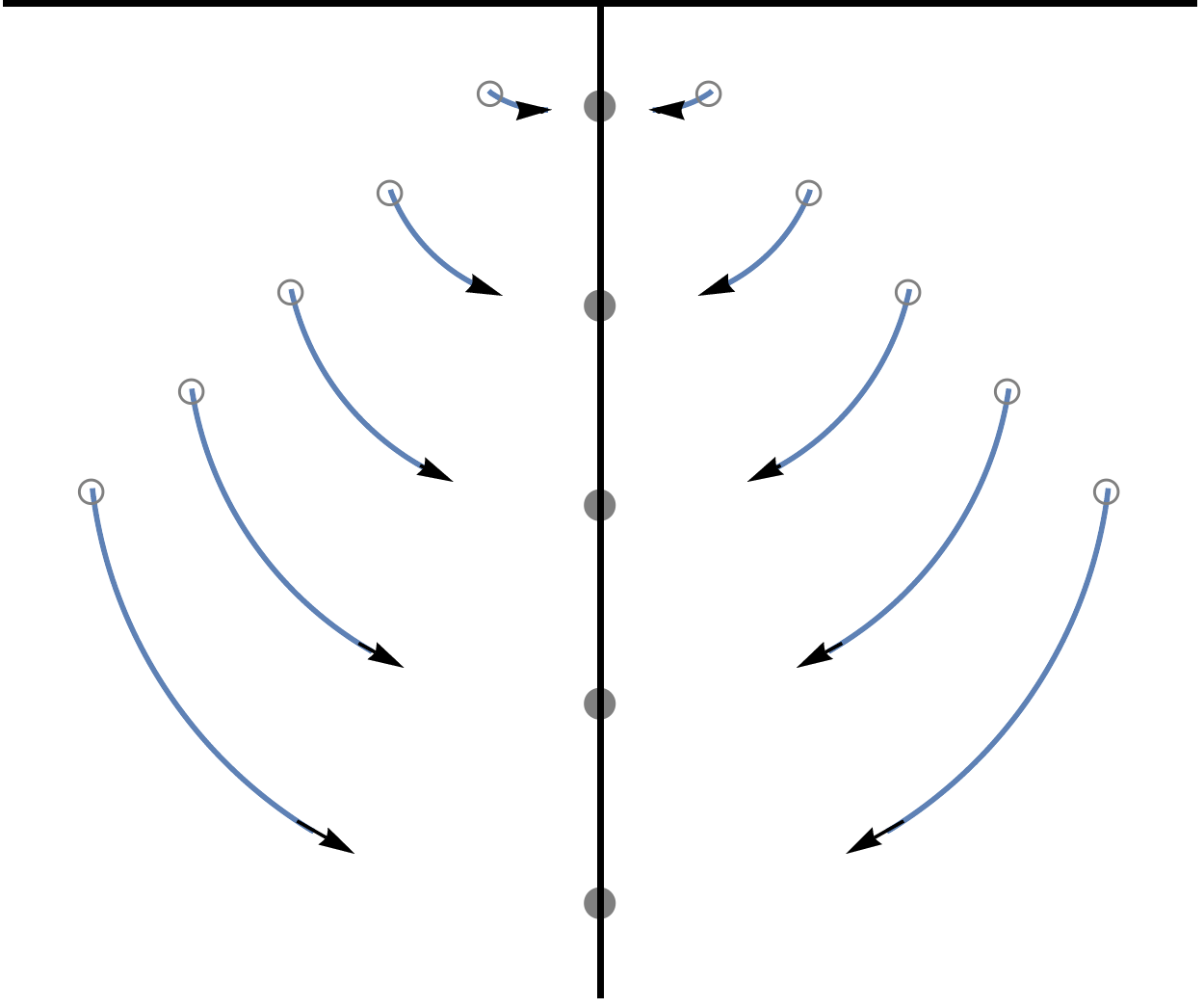}
 		\put (90,78) {$\displaystyle\omega_{\text{Re}}$}
		\put (35,3) {$\displaystyle-\omega_{\text{Im}}$}
		
		\put (93,38) {$\displaystyle z=1$}
		\put (84,20) {$\displaystyle z>1$}
		\put (52,8.5) {$\displaystyle z\geq d-1$}
		
		\put (26,78) {$\displaystyle n=0$}
		\put (17,69) {$\displaystyle n=1$}
		\put (7,60) {$\displaystyle n=2$}
		\put (0,51) {$\displaystyle etc.$}

		\end{overpic}
        \end{subfigure}
        \caption{The hollow dots denote the location of the quasinormal modes when $z=1$\,. The $n$ denotes the overtone number. When increasing $z$, the location of quasinormal mode will follow the arc towards the vertical axis. At $z=d-1$ the mode, for the first time, hits the vertical axis. From there on, for any $z\geq d-1$\,, the quasinormal mode remains somewhere on the vertical axis. We stress that remaining on the vertical axis is because of the vanishing real part when $z\geq d-1$\, and corresponds to overdamped systems.}\label{figure000}
\end{figure}
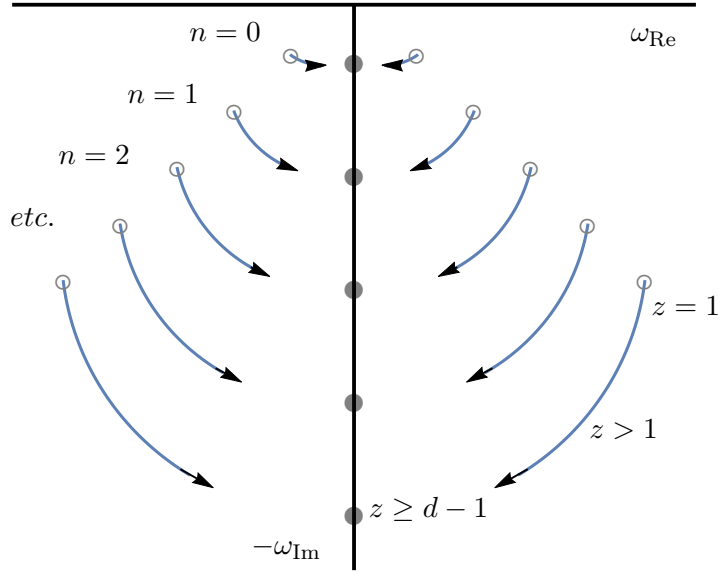

We focus on $d=2, 3, 4$ because of their relevance in real world systems. We choose momentum to be vanishing and leave the $k\neq0$ to future work. For clearness we restrict ourselves to $\Delta=3, 4, 5.5$ and values of $z$ for 1 through 6. To get some intuition for the behavior of the quasinormal modes in the complex plane, when varying $z$\,, we present a cartoon in Figure \ref{figure000}.

\begin{figure}[h]
        \centering
      \begin{subfigure}[b]{.45\textwidth}
                \begin{overpic}[width=1\textwidth]{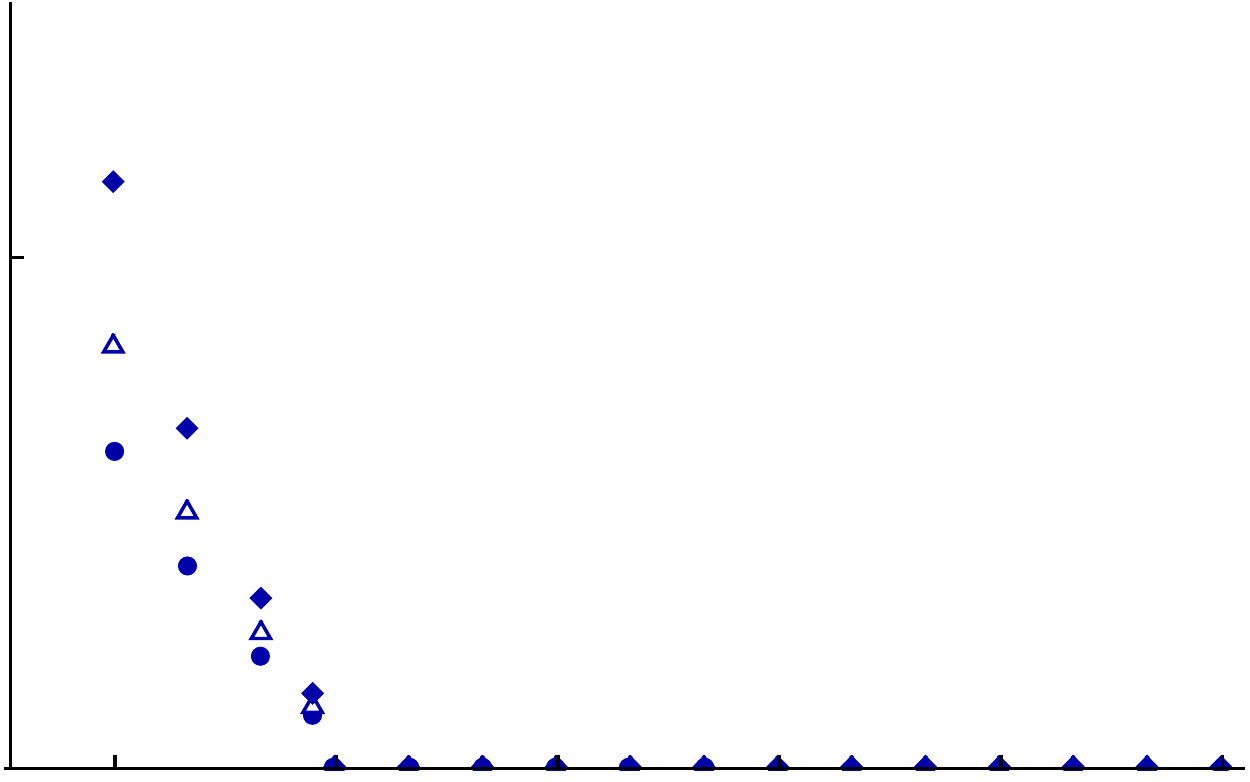}
 		\put (-17.2,56) {$\displaystyle\frac{\text{Re}(\omega_{0})}{4\pi T}$}
		\put (99,3) {$\displaystyle z$}
		
		\put (45,60) {$\displaystyle \underline{\bf{d=3}}$}
		
		\put (25.,63) {\rotatebox{-90}{$\displaystyle -----------$}}
		\put (28,57) {\rotatebox{-90}{$\displaystyle d=z+1$}}
		
		\put (-3,0) {$\displaystyle 0$}
		\put (-3,40.5) {$\displaystyle 1$}
		
		\put (8,-3) {$\displaystyle 1$}
		\put (25.5,-3) {$\displaystyle 2$}
		\put (43.5,-3) {$\displaystyle 3$}
		\put (61,-3) {$\displaystyle 4$}
		\put (78.5,-3) {$\displaystyle 5$}
		\put (96.5,-3) {$\displaystyle 6$}	
		\end{overpic}
        \end{subfigure}
        \qquad
        \begin{subfigure}[b]{.45\textwidth}
                \begin{overpic}[width=1\textwidth]{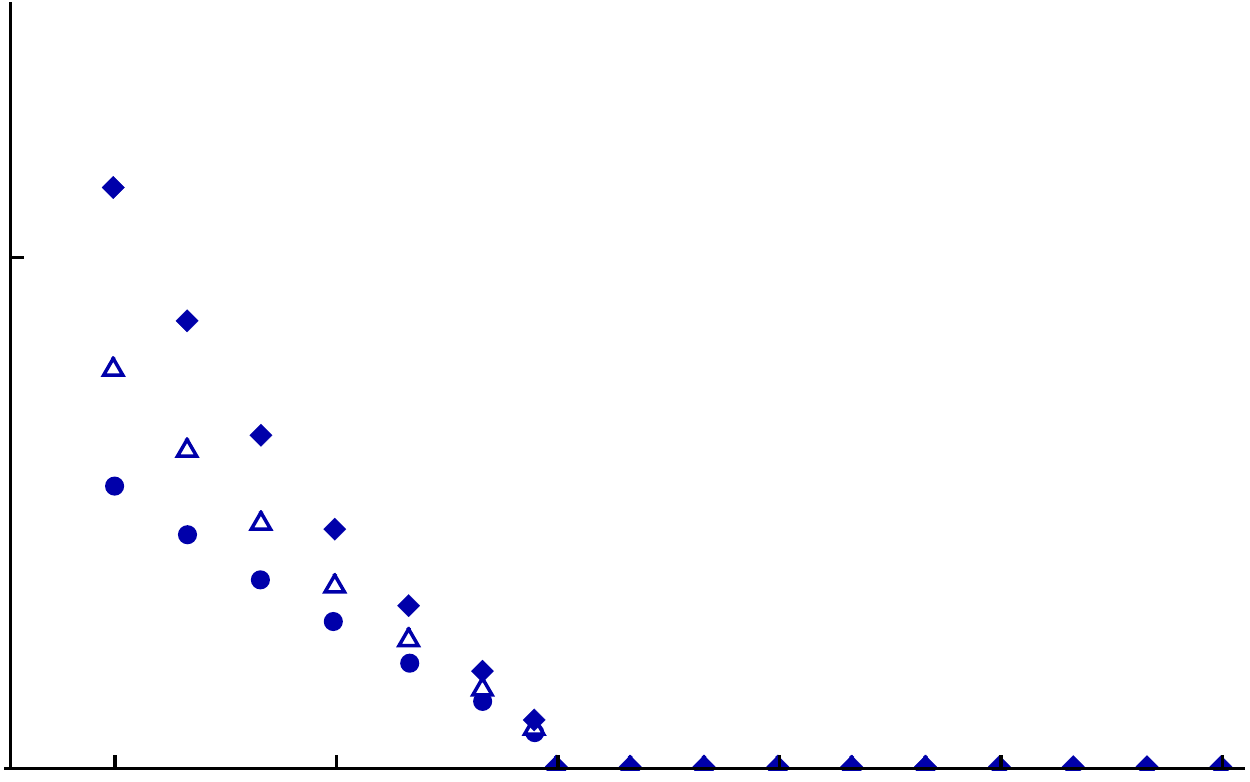}
 		\put (-17.2,56) {$\displaystyle\frac{\text{Re}(\omega_{0})}{4\pi T}$}
		\put (99,3) {$\displaystyle z$}
		
		\put (45,60) {$\displaystyle \underline{\bf{d=4}}$}
		
		\put (42.7,63) {\rotatebox{-90}{$\displaystyle -----------$}}
		\put (45.5,57) {\rotatebox{-90}{$\displaystyle d=z+1$}}
		
		\put (-3,0) {$\displaystyle 0$}
		\put (-3,40.5) {$\displaystyle 1$}
		
		\put (8,-3) {$\displaystyle 1$}
		\put (25.5,-3) {$\displaystyle 2$}
		\put (43.5,-3) {$\displaystyle 3$}
		\put (61,-3) {$\displaystyle 4$}
		\put (78.5,-3) {$\displaystyle 5$}
		\put (96.5,-3) {$\displaystyle 6$}		
		\end{overpic}
        \end{subfigure}
        \caption{The real part of the computed quasinormal modes. For $d=2$ the value is zero everywhere, as can be read off from Table 2 in the appendix. Diamonds correspond to $\Delta=5.5$, triangles correspond to $\Delta=4$ and the dots correspond to $\Delta=3$\,.}\label{figureB}
\end{figure}
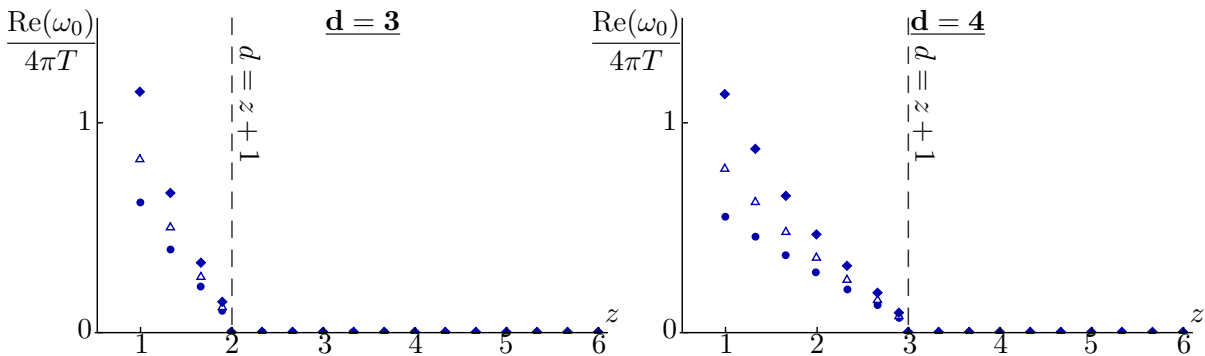

\begin{figure}[h]
        \centering
        \begin{subfigure}[b]{.60\textwidth}
                \begin{overpic}[width=1\textwidth]{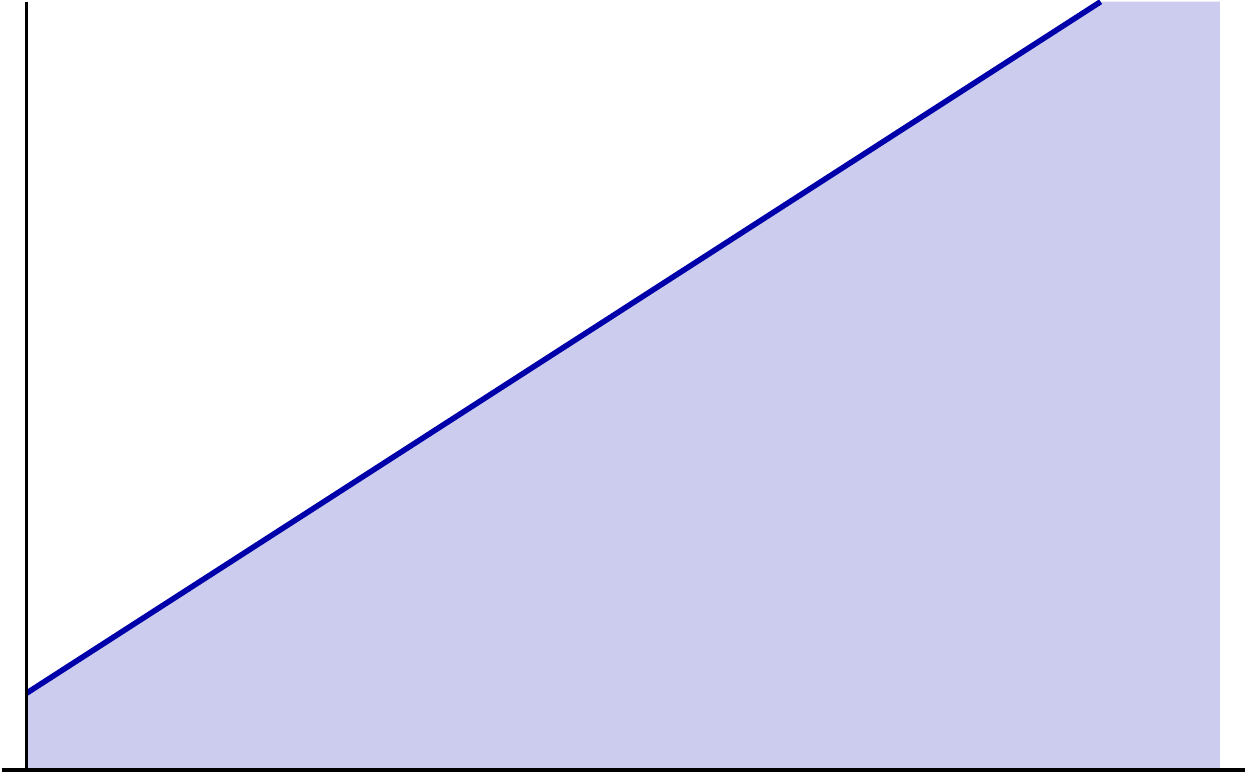}
 		\put (-3,59) {$\displaystyle d$}
		\put (12,32) {$\displaystyle -\omega_{\text{Im}}$}
		\put (40,54) {$\displaystyle \omega_{\text{Re}}$}
		
		\put (58,7) {$\displaystyle -\omega_{\text{Im}}$}
		\put (87,29) {$\displaystyle \omega_{\text{Re}}$}
		
		\put (10,60) {$\displaystyle \text{non-overdamped}$}
		\put (100,1) {$\displaystyle z$}
		\put (3,30) {$\displaystyle \begin{overpic}[width=0.45\textwidth]{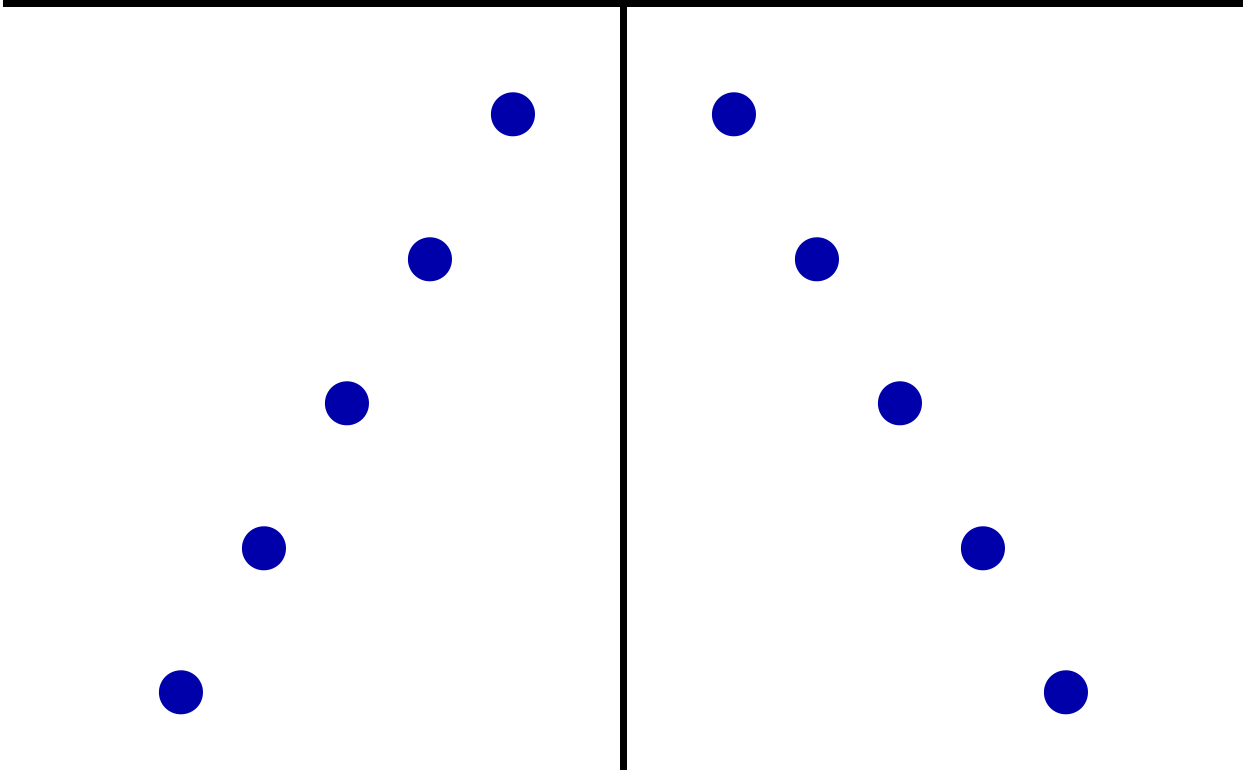}\end{overpic}$}
		\put (61,48) {\rotatebox{32}{$\displaystyle d=z+1$}}
		\put (59,35) {$\displaystyle \text{overdamped}$}
		\put (50,05) {$\displaystyle \begin{overpic}[width=0.45\textwidth]{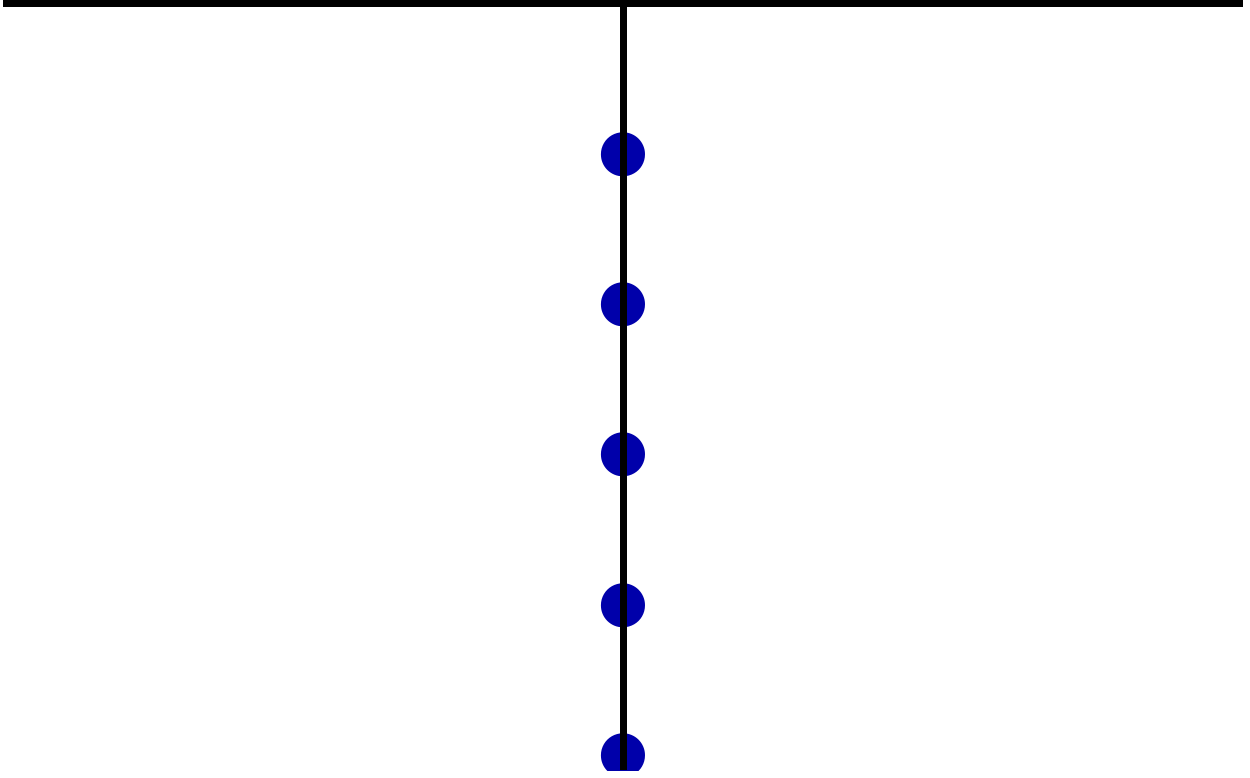}\end{overpic}$}
		\end{overpic}
        \end{subfigure}
        \caption{The line given by $d=z+1$ divides quasinormal modes in being overdamped or non-overdamped. Both plots within the separated regions show the qualitative behavior of the overtones of the quasinormal modes in those regions. Exactly at and underneath the line $d=z+1$ the system is overdamped.}\label{figure00}
\end{figure}
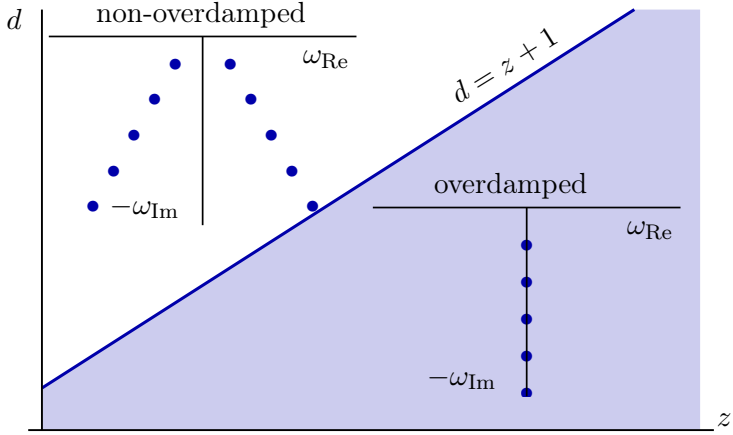

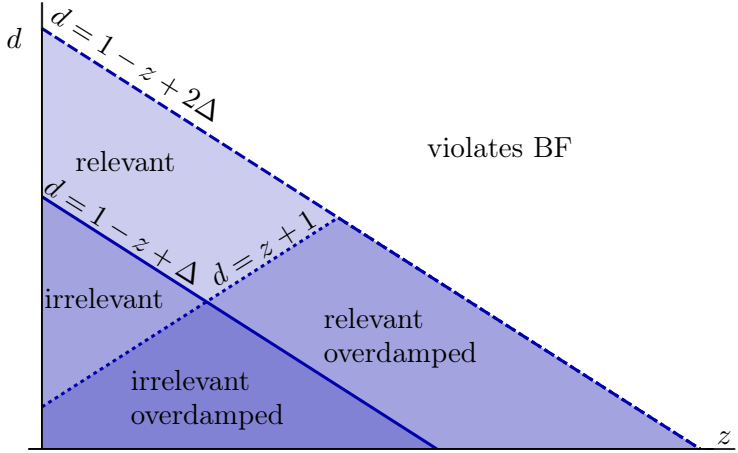
\begin{figure}[h]
        \centering
        \vspace{0.4cm}
                \begin{subfigure}[b]{.60\textwidth}
                \begin{overpic}[width=1\textwidth]{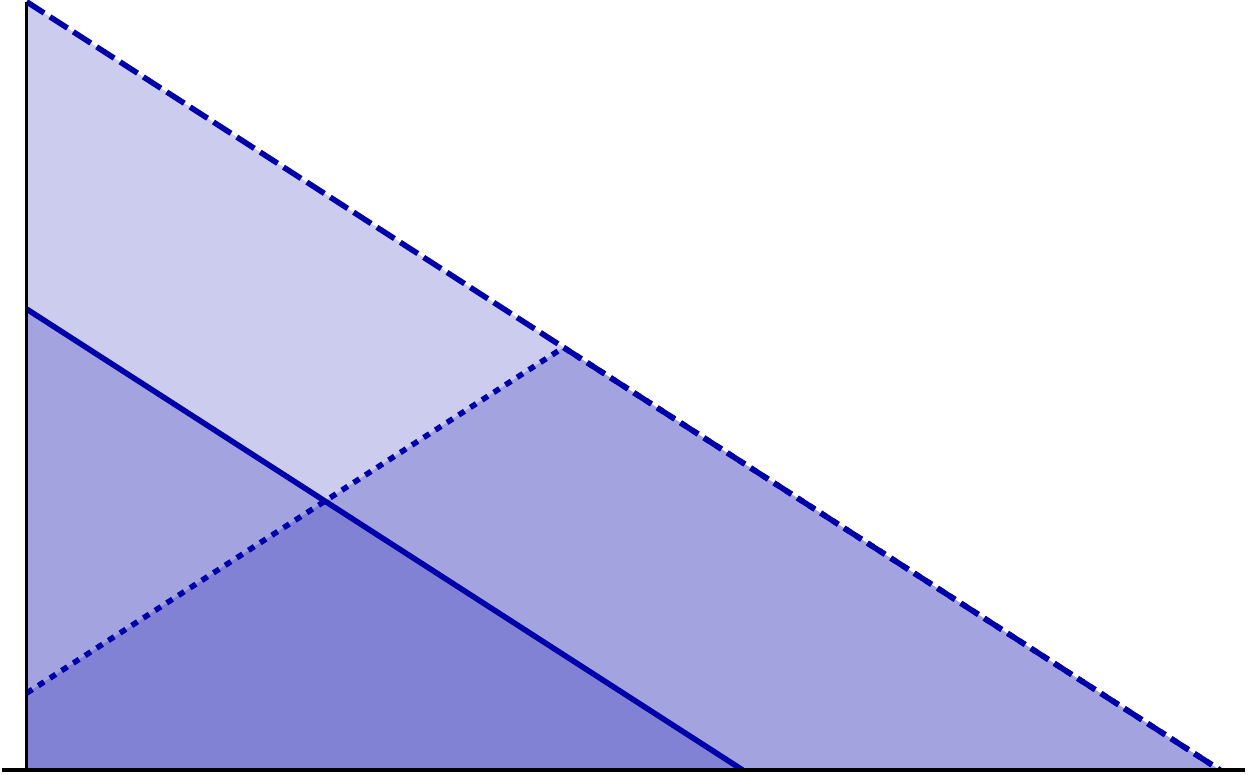}

                	\put (+1.64,62.5) {|}
		\put (+1.53,62.5) {|}
		\put (98.5,-0.18){\rotatebox{90}{|}}
		\put (98.5,-0.28853){\rotatebox{90}{|}}

 		\put (-3,59) {$d$}
		\put (7,41) {$\displaystyle \text{relevant}$}
		\put (2.5,21) {$\displaystyle \text{irrelevant}$}
		\put (100,1) {$\displaystyle z$}
		\put (26,24) {\rotatebox{32}{$\displaystyle d=z+1$}}
		\put (58,43) {$\displaystyle \text{violates BF}$}
		\put (43,18) {$\displaystyle \text{relevant}$}
		\put (43,13) {$\displaystyle \text{overdamped}$}
		\put (15,9) {$\displaystyle \text{irrelevant}$}
		\put (15,4) {$\displaystyle \text{overdamped}$}
		\put (2,38) {\rotatebox{-32}{$\displaystyle d=1-z+\Delta$}}
		\put (3,63){\rotatebox{-32}{$\displaystyle d=1-z+2\Delta$}}
		\end{overpic}
        \end{subfigure}
        \caption{The information from Figures \ref{figure0} and \ref{figure00} is summarized. Exactly at the $d=1-z+\Delta$ line the operator is marginal. On and above the line $d=1-z+2\Delta$ the BF bound is violated. Exactly at and underneath the line $d=z+1$ the system is overdamped.}\label{figureA}
\end{figure}

In Figure \ref{figureB} we plot the real part of the quasinormal modes versus $z$. A sample of the data can also be found in Table \ref{table2} in the appendix.
In the region $d\leq z+1$ we find that the system is overdamped. For $d>z+1$ we find a non-zero real part. This result is interesting when taking into account the conjectures of \cite{Abdalla:2012si, Myung:2012cb}\,, which state that for (most) Lifshitz black holes the quasinormal modes are purely imaginary. Moreover, finding non-overdamped cases for $d>z+1$ contrasts the overdamped cases found for $d\geq4$\,, $z=2$ in a $R^{2}$ gravity setting \cite{Abdalla:2012si} and $d\geq2$\,, $z=2$ in a $R^{3}$ gravity setting \cite{Giacomini:2012aa}. We present a cartoon of the qualitative structure of the overtones of the quasinormal modes in the complex plane in Figure \ref{figure00}.
We summarize Figures \ref{figure0} and \ref{figure00} in Figure \ref{figureA} for clarity.
Notice that for fixed $\Delta$\,, $d$ and increasing $z$\,, the real part is strictly decreasing, until it hits $d=z+1$\,. When fixing $z$ we have the relation that higher $\Delta$ corresponds to a greater real part in the non-overdamped region. 
\section{Relaxation times}
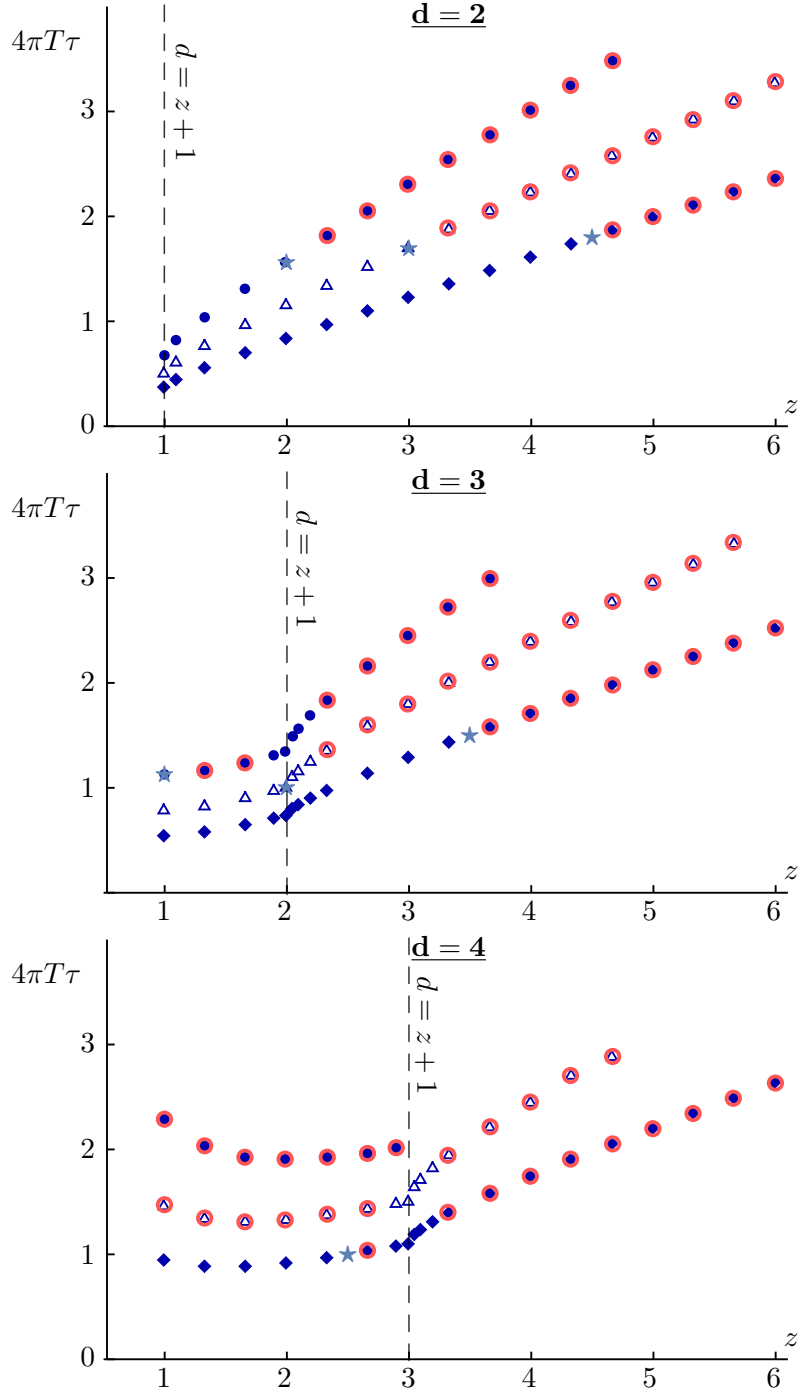
\begin{figure}[h]
        \centering
        \begin{subfigure}[b]{.60\textwidth}
                \begin{overpic}[width=1\textwidth]{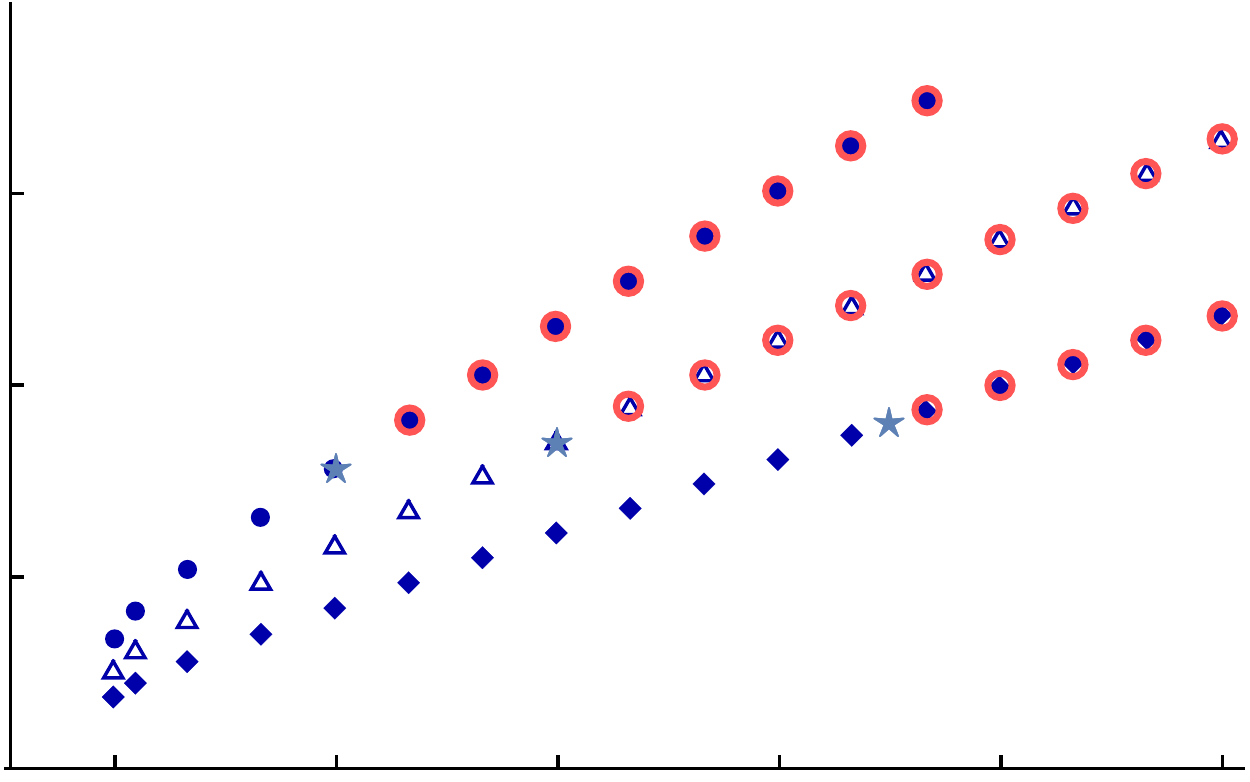}
 		\put (-13,56) {$\displaystyle4\pi T \tau$}
		\put (99,3) {$\displaystyle z$}
		
		\put (45,60) {$\displaystyle \underline{\bf{d=2}}$}
		
		\put (7.75,63) {\rotatebox{-90}{$\displaystyle ---------------$}}
		\put (10.5,57) {\rotatebox{-90}{$\displaystyle d=z+1$}}
		
		\put (-3,0) {$\displaystyle 0$}
		\put (-3,15) {$\displaystyle 1$}
		\put (-3,30.5) {$\displaystyle 2$}
		\put (-3,46) {$\displaystyle 3$}
		
		\put (8,-3) {$\displaystyle 1$}
		\put (25.5,-3) {$\displaystyle 2$}
		\put (43.5,-3) {$\displaystyle 3$}
		\put (61,-3) {$\displaystyle 4$}
		\put (78.5,-3) {$\displaystyle 5$}
		\put (96.5,-3) {$\displaystyle 6$}
		
		\end{overpic}
        \end{subfigure}
       		 \\
        \hfill
        \\
              \begin{subfigure}[b]{.60\textwidth}
                \begin{overpic}[width=1\textwidth]{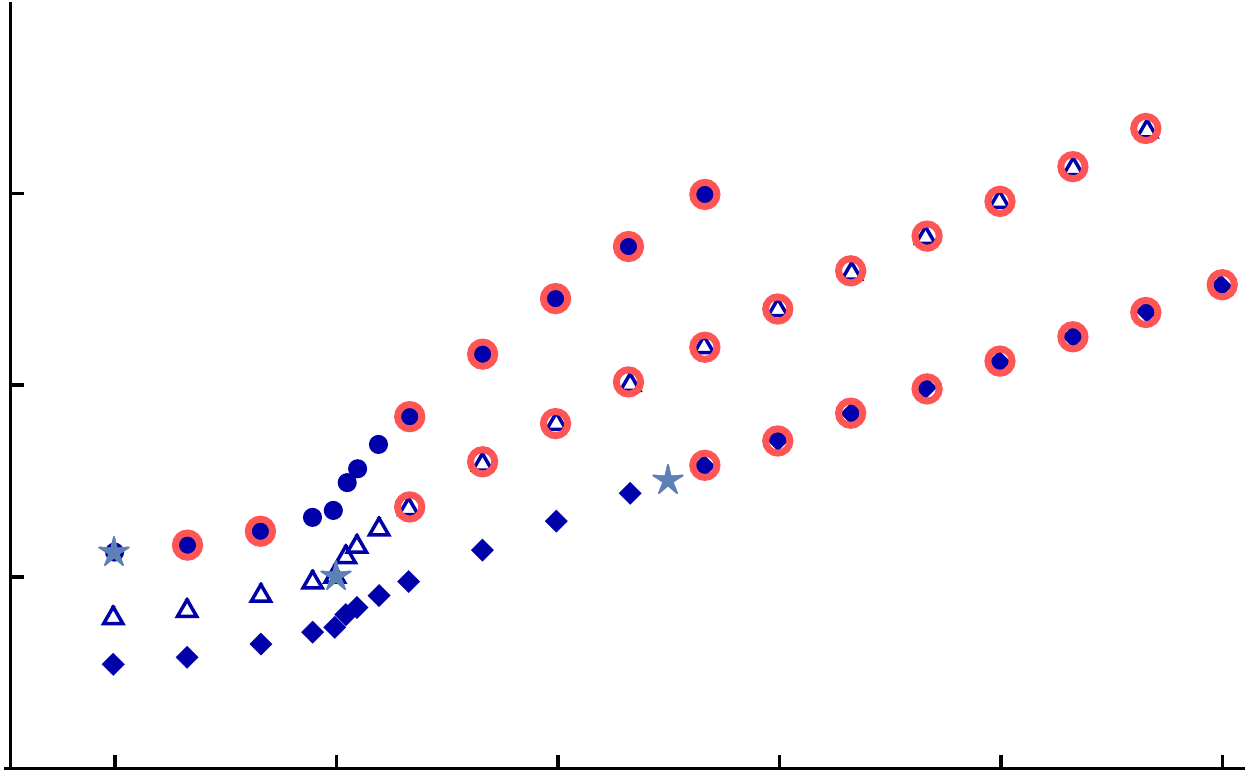}
 		\put (-13,56) {$\displaystyle4\pi T \tau$}
		\put (99,3) {$\displaystyle z$}
		
		\put (25.55,63) {\rotatebox{-90}{$\displaystyle ---------------$}}
		\put (28,57) {\rotatebox{-90}{$\displaystyle d=z+1$}}
		
		\put (45,60) {$\displaystyle \underline{\bf{d=3}}$}
		
		\put (-3,15) {$\displaystyle 1$}
		\put (-3,30.5) {$\displaystyle 2$}
		\put (-3,46) {$\displaystyle 3$}
		
		\put (8,-3) {$\displaystyle 1$}
		\put (25.5,-3) {$\displaystyle 2$}
		\put (43.5,-3) {$\displaystyle 3$}
		\put (61,-3) {$\displaystyle 4$}
		\put (78.5,-3) {$\displaystyle 5$}
		\put (96.5,-3) {$\displaystyle 6$}
		\end{overpic}
        \end{subfigure}
        \\
        \hfill
        \\
        \begin{subfigure}[b]{.60\textwidth}
                \begin{overpic}[width=1\textwidth]{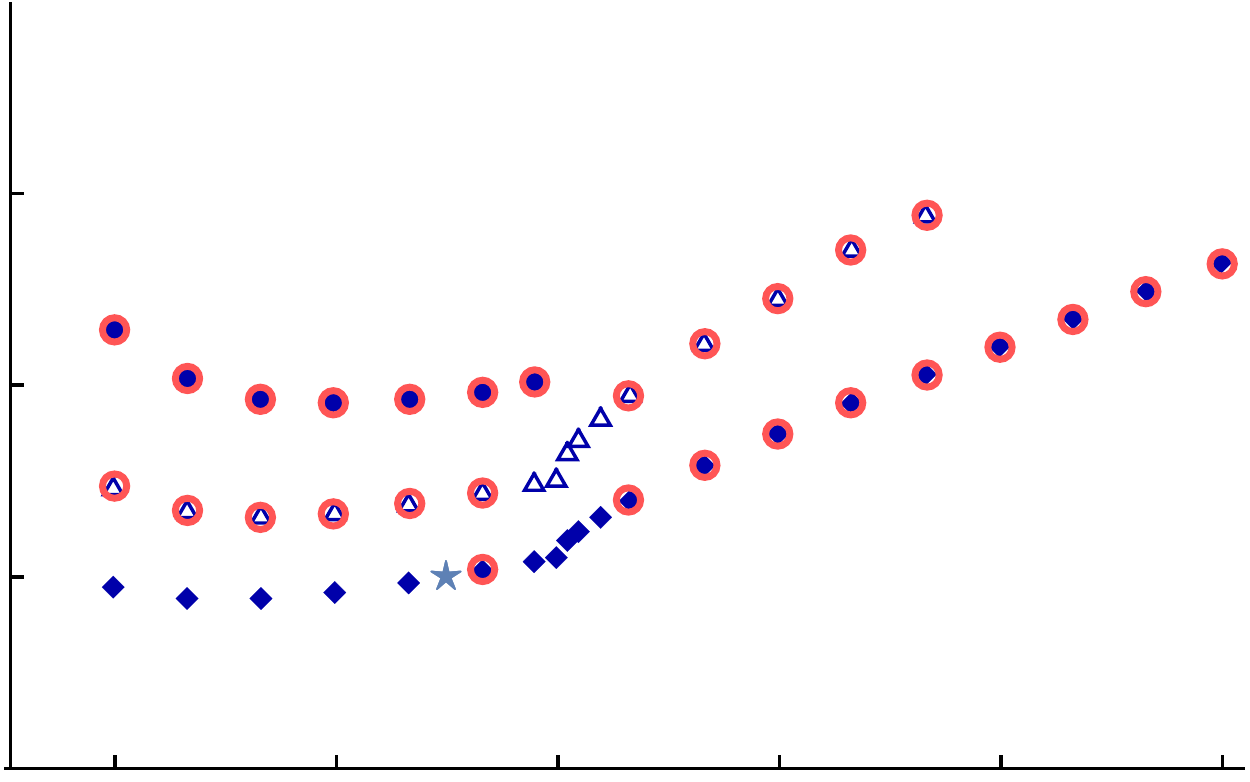}
 		\put (-13,56) {$\displaystyle4\pi T \tau$}
		\put (99,3) {$\displaystyle z$}
		
		\put (45,60) {$\displaystyle \underline{\bf{d=4}}$}
		
		\put (43.25,63) {\rotatebox{-90}{$\displaystyle ---------------$}}
		\put (45.5,57) {\rotatebox{-90}{$\displaystyle d=z+1$}}
		
		\put (-3,0) {$\displaystyle 0$}
		\put (-3,15) {$\displaystyle 1$}
		\put (-3,30.5) {$\displaystyle 2$}
		\put (-3,46) {$\displaystyle 3$}
		
		\put (8,-3) {$\displaystyle 1$}
		\put (25.5,-3) {$\displaystyle 2$}
		\put (43.5,-3) {$\displaystyle 3$}
		\put (61,-3) {$\displaystyle 4$}
		\put (78.5,-3) {$\displaystyle 5$}
		\put (96.5,-3) {$\displaystyle 6$}
		
		\end{overpic}
        \end{subfigure}
        \caption{Results of the relaxation times $\tau$\,, numerically computed from quasinormal modes. Diamonds correspond to $\Delta=5.5$\,, triangles correspond to $\Delta=4$ and the dots correspond to $\Delta=3$\,. A star denotes that the operator at that point is marginal. The curve stops on the right when the BF bound is violated. The highlighted points denote when the operator is relevant, otherwise the point is irrelevant. However, around the region $d=z+1$ we did not put any highlights for readability's sake.}\label{figureC}
\end{figure}

In Figure \ref{figureC} we present the relaxation time versus $z$\,.
For the relaxation times the point $d=z+1$ leaves its footprint as well as in the real part, by separating different behaviors on either side of this point. A sample of the data points is given in Table 2 in the appendix. Features which we gather from Figure \ref{figureC} are:  
\begin{itemize}
    \item Higher scaling dimension $\Delta$ of an operator corresponds to a lower relaxation time.
    \item Increasing $z$\,, the amount of anisotropy, corresponds to a higher relaxation time $\tau$ when $d\leq z+1$\,.
    \item Overdampedness is independent of the dimension of the operator.
\end{itemize}
\section{Outlook}
The results in this paper were obtained from the holographic point of view. It would be interesting if, in the spirit of \cite{Birmingham:2001pj}, one could reproduce these results directly from the field theory side. In particular it would be interesting to understand the difference between the regimes $d\leq z+1$ and $d>z+1$. In context of this we make the following observation. Consider a free $d$-dimensional theory with Lifshitz scaling and a dispersion
\begin{equation}
	\omega
	\sim
	|k|^{z}
	\,.
\end{equation}
The number of states $\Omega(k)$ up to momentum $|k|$ is obtained by taking a spherical $k$-space volume and dividing it by the volume occupied per allowed state, resulting in
\begin{equation}
	\Omega(k)
	\sim
	k^{d-1}
	\,.
\end{equation}
The density of states $D(\omega)$ is defined as
\begin{equation}
	D(\omega)
	=
	\frac{d\Omega(k(\omega))}{d\omega}
	=
	\frac{d\Omega(k)}{dk}
	\frac{dk(\omega)}{d\omega}
	\sim
	\omega^{\frac{d-2}{z}}
	\omega^{\frac{1-z}{z}}
	=
	\omega^{\frac{d-(z+1)}{z}}
	\,.
\end{equation}
Notice that there is a qualitatively different behavior between the regimes $d<z+1$ and $d>z+1$. For $d<z+1$, the density of states decreases with energy, whereas for $d>z+1$ it increases with energy. Perhaps this behavior is related to the non-overdamped and overdamped phases after adding interactions. We leave this for future research.
\acknowledgments
It is a privilege to thank Vivian Jacobs and Henk Stoof for interesting and stimulating discussions. This work was supported by the Netherlands Organisation for Scientific Research (NWO) under the VICI grant 680-47-603, and the Delta-Institute for Theoretical Physics (D-ITP) that is funded by the Dutch Ministry of Education, Culture and Science (OCW).

\bibliography{/Users/Watse/Dropbox/Universiteit/Bibtex/bib2}
\appendix
\setcounter{secnumdepth}{0}
\section{Appendix}
In Table \ref{table1} we present data in order to compare to the results of \cite{Horowitz:2000aa}. Table \ref{table2} contains a sample of the results used in this paper.
\begin{table}[h]
\center
	\tabcolsep=0.11cm
	\begin{tabular}{ |c|cc|cc|cc| }
		\hline
		&\multicolumn{2}{|c|}{$d=3$}&\multicolumn{2}{|c|}{$d=4$}&\multicolumn{2}{|c|}{$d=6$}\\
		\hline
		$r_{h}$&$\omega_{Im}$&$\omega_{Re}$&$\omega_{Im}$&$\omega_{Re}$&$\omega_{Im}$&$\omega_{Re}$\\
		\hline
		  100 & 266.38  & 184.94  & 274.66 	& 311.94 	& 261.24 & 500.74 \\
		  50 	& 133.19 	& 92.47 	& 137.33 	& 155.97 	& 130.62 & 250.37\\
		  10	& 26.63 	& 18.49 	& 27.46 	& 31.19 	& 26.12 & 50.07\\
		  5 	& 13.31 	& 9.24 	& 13.73 	& 15.59 	& 13.06 & 25.03\\
		  1 	& 2.66 	& 1.84 	& 2.74 	& 3.11 	&  2.61& 5.00\\
		\hline
	\end{tabular}
	\caption{Results which can be compared to \cite{Horowitz:2000aa}. Notice that this corresponds to putting the mass $m$ to zero and $z=1$.}\label{table1}
\end{table}

\begin{table}[h]
\center
\tabcolsep=0.11cm 
\begin{tabular}{ |c|cc|cc|cc|cc|cc|cc| }
\hline
  &\multicolumn{6}{|c|}{$\Delta=3$}
  &\multicolumn{6}{|c|}{$\Delta=5.5$}
  \\
  \hline
  &\multicolumn{2}{|c|}{$d=2$}&\multicolumn{2}{|c|}{$d=3$}&\multicolumn{2}{|c|}{$d=4$}
  &\multicolumn{2}{|c|}{$d=2$}&\multicolumn{2}{|c|}{$d=3$}&\multicolumn{2}{|c|}{$d=4$} \\
  \hline
  $z$ & $\frac{\text{Re}(\omega_{0})}{4\pi T}$ &$4\pi T\tau $& $\frac{\text{Re}(\omega_{0})}{4\pi T}$ &$4\pi T\tau $& $\frac{\text{Re}(\omega_{0})}{4\pi T}$ &$4\pi T\tau $
  & $\frac{\text{Re}(\omega_{0})}{4\pi T}$ &$4\pi T\tau $& $\frac{\text{Re}(\omega_{0})}{4\pi T}$ &$4\pi T\tau $& $\frac{\text{Re}(\omega_{0})}{4\pi T}$ &$4\pi T\tau $  \\
  \hline
  1 	& 0	  	& 2/3   & 0.61 & 1.12 & 0.54  & 2.27 &0       & 4/11& 1.14& 0.53& 1.13& 0.93\\
  4/3 	& 0.00 	& 1.03 & 0.39 & 1.15 & 0.45  & 2.01 &0.00  & 0.54& 0.66& 0.57& 0.87& 0.87\\
  5/3	& 0.00 	& 1.30 & 0.21 & 1.23 & 0.36  & 1.91 &0.00  & 0.69& 0.32& 0.63& 0.64& 0.87\\
  2	& 0.00 	& 1.55 & 0      & 4/3   & 0.28  & 1.89 &0.00  & 0.82& 0     & 8/11& 0.46& 0.90\\
  7/3	& 0.00 	& 1.80 & 0.00 & 1.83 & 0.204& 1.90 &0.00  & 0.95& 0.00& 0.96& 0.31& 0.95\\
  8/3	& 0.00 	& 2.04 & 0.00 & 2.15 & 0.13  & 1.94 &0.00  & 1.09& 0.00& 1.13& 0.18& 1.02\\
  3	& 0.00 	& 2.28 & 0.00 & 2.44 & 	     &		&0.00  & 1.21& 0.00& 1.28& 0     & 12/11\\
  10/3& 0.00 	& 2.52 & 0.00 & 2.72 &          &		&0.00  & 1.34& 0.00& 1.42& 0.00& 1.38\\
  11/3& 0.00 	& 2.76 & 0.00 & 2.99 &          &		&0.00  & 1.47& 0.00& 1.56& 0.00& 1.57\\
  4	& 0.00 	& 2.99 &         &         &     	     &		&0.00  & 1.60& 0.00& 1.70& 0.00& 1.73\\
  13/3& 0.00 	& 3.23 &         &         &          &		&0.00  & 1.72& 0.00& 1.84& 0.00& 1.89\\
  14/3& 0.00 	& 3.46 &         &         &          &		&0.00  & 1.85& 0.00& 1.97& 0.00& 2.04\\
  5	& 	 	&	   &         & 	 &          &		&0.00  & 1.98& 0.00& 2.11& 0.00& 2.19\\
  16/3& 		&  	   &         & 	 &          &		&0.00  & 2.10& 0.00& 2.24& 0.00& 2.33\\
  17/3&	 	&         &         & 	 &          &		&0.00  & 2.23& 0.00& 2.37& 0.00& 2.48\\
  6	&  		&         &         & 	 &          &		&0.00  & 2.35& 0.00& 2.50& 0.00& 2.63\\
  \hline
\end{tabular}
\caption{Parts of the results which are used to plot Figure \ref{figureB} and \ref{figureC}. Empty rows signal violation of the BF bound.}\label{table2}
\end{table}

\end{document}